\documentclass[aps,prx,reprint, amsmath, amssymb,superscriptaddress]{revtex4-1}

\usepackage{lmodern}
\usepackage[utf8]{inputenc}

\usepackage{bm}
\usepackage[retainorgcmds]{IEEEtrantools}
\usepackage{graphicx}
\usepackage{mathrsfs}
\usepackage{amsmath}
\usepackage{amssymb}
\usepackage{color}
\usepackage{amsfonts}
\usepackage{times,txfonts}
\usepackage{nicefrac}
\usepackage[colorlinks=true,linkcolor=blue,urlcolor=blue,citecolor=blue,pdfusetitle]{hyperref}
\usepackage{physics}
\usepackage{soul}
\usepackage{dsfont}
\usepackage{geometry}
 \usepackage[dvipsnames]{xcolor}
 \usepackage[authormarkup=none]{changes}
 \definechangesauthor[name={Adalberto}, color=red]{g}

\newcommand{\sigB}{\Lambda_\text{cl}}
\newcommand{\sigA}{\Lambda_\text{qu}}

 \newcommand{\sigBij}{\lambda_\text{cl}}
 \newcommand{\sigAij}{\lambda_\text{qu}}

\newcommand{\maxmix}{\frac{\mathds{1}}{d}}

\begin{document}

\title{Quantum quench thermodynamics at high temperatures}
\date{\today}
\author{Adalberto D. Varizi}
    \affiliation{Departamento de F\'isica, Instituto de Ci\^encias Exatas, Universidade Federal de Minas Gerais, 30123-970, Belo Horizonte, Minas Gerais, Brazil}
\author{Raphael C. Drumond}
    \affiliation{Departamento de Matem\'atica, Instituto de Ci\^encias Exatas, Universidade Federal de Minas Gerais, 30123-970, Belo Horizonte, Minas Gerais, Brazil}
\author{Gabriel T. Landi}
    \affiliation{Instituto de F\'isica da Universidade de S\~ao Paulo,  05314-970 S\~ao Paulo, Brazil}

\begin{abstract}
The entropy produced when a system undergoes an infinitesimal quench is directly linked to the work parameter susceptibility, making it sensitive to the existence of a quantum critical point.
Its singular behavior at $T=0$, however,  disappears as the temperature is raised, hindering its use as a tool for spotting quantum phase transitions.
Notwithstanding the entropy production can be split into  classical and quantum components, related with changes in populations and coherences.
In this paper we show that these individual contributions continue to exhibit signatures of the quantum phase transition, even at arbitrarily high temperatures.
This is a consequence of their intrinsic connection to the derivatives of the energy eigenvalues and eigenbasis.
We illustrate our results in two prototypical quantum critical systems, the Landau-Zener and $XY$ models.
\end{abstract}

\maketitle{}

%
%
\section{\label{sec:int}Introduction}
At zero temperature a quantum many-body system can manifest distinct behaviors depending on the values of its internal couplings.
And, as these values are altered, quantum fluctuations may drive the system through a phase transition.
These quantum phase transitions are characterized by stark modifications in the system's ground-state, frequently associated with a symmetry breaking at a critical point.

Such transitions can be characterized with a variety of information theoretic tools, such as the ground-state fidelity, fidelity susceptibility and Loschmidt echo~\cite{Gu2010}.
For finite systems, a large overlap between the ground states of nearby Hamiltonians in parameter space is expected, giving a fidelity close to unity.
In contrast, the distinct properties of the ground states pertaining to different phases make them more ``distant", and this is revealed as a sharp drop in fidelity in the vicinity of the transition point~\cite{Zanardi2006}. 
Equivalently, this effect can be verified as an increase in the fidelity rate change, the fidelity susceptibility, signaling the existence of a phase transition~\cite{Zanardi2006,Gu2010}.

In a similar way, the Loschmidt echo, which behaves as a type of dynamic fidelity, gives a measure of the distinguishability between the time-evolved ground states of an unperturbed Hamiltonian and its perturbed counterpart.
When the system is quenched through a critical point, the echo may have a dip at the critical parameter and a series of decays and revivals as a function of time, also indicating the presence of critical behavior~\cite{Quan2006,Heyl2013}.

A shortcoming of these approaches is that they are all based on the zero temperature ground states.
As the temperature $T$ is raised above zero, a competition between quantum and thermal fluctuations emerge, with the latter quickly dominating the physics of the problem~\cite{Sachdev2011}.
Nonetheless,  fingerprints of the quantum critical point are still visible for finite temperatures, albeit not as sharply as when $T=0$.

A generalization of the ground state fidelity and Loschmidt echo approaches to thermal states was put forward in~\cite{Zanardi2007}.
A decay in their values close to the quantum critical point was also observed, although gradually less sharp with rising temperatures, up to a complete disappearance of any distinct feature.

On another front, the correlations and/or coherences in the system have been likewise used as tools for locating transition points.
This started with entanglement measures~\cite{Osterloh2002,Wu2004} (or some derivatives of it), which can show nonanalytic behavior at a quantum critical point.
Similar analysis was later extended to mutual information, classical correlations and quantum discord~\cite{Dillenschneider2008,Sarandy2009}, and more recently to quantifiers of coherence~\cite{Karpat2014,Malvezzi2016,Chen2016coh,LiLin2016,Lei2016,Yi2019,Hu2021} in several spin models.
Commonly, the singularities in these quantities disappear at finite $T$, and attempts to estimate the critical point from some extrema of these functions generally becomes poorer with increasing temperature.

An exception, though, was demonstrated in~\cite{Werlang2010} for quantum discord.
In this case, the authors showed that a kink in the discord in the state of two nearest neighbors spins, reduced from a global thermal state, in the XXZ model with no external field, indicates the quantum critical points in the system even at high temperatures.
Still, further analyses with other models~\cite{Werlang2011,LiLin2011} showed that for some transitions these nonanalyticities can, again, disappear, with the peaks or valleys that replace them becoming less pronounced and displaced from the true critical point with increasing temperature. 

Quantum phase transitions have also played a similar role in the context of  quantum thermodynamics, particular in the case of unitary work protocols. In fact, in Ref.~\cite{Dorner2012,Mascarenhas2014} it was shown that a divergence in the entropy
production during a sudden quench pinpointed the existence of a quantum critical point.
The entropy production is generally written as a quantum relative entropy, and, therefore, similarly gives a measure of distinguishability between states.
Here, a temperature is naturally introduced from the start, but again, the divergence at $T=0$ is smoothed as $T$ increases~\cite{Dorner2012}.

The entropy production, however, can be consistently split into two parts, associated with classical and quantum contributions, the latter steaming from quantum coherences~\cite{Santos2019,Francica2019,Scandi2019,Varizi2021}.
Beyond unitary work protocols~\cite{Francica2019}, particularly in critical systems~\cite{Varizi2020,Varizi2021}, this type of splittings has found many applications like in work extraction protocols~\cite{Francica2020}, relaxation towards equilibrium~\cite{Santos2019,Mohammady2020}, quasi-static evolution and consequences of coherences to the fluctuation-dissipation theorem~\cite{Miller2019,Scandi2019}, thermodynamics resource theory~\cite{Lostaglio2015} and quantum optics thermodynamics~\cite{Elouard2020}.
In~\cite{Varizi2020,Varizi2021} we hinted at the fact that this contributions could individually signal the existence of a critical point at \emph{any temperature}.
The aim of this paper is to further explore and clarify this point.

Hence, we consider a system with Hamiltonian $H(g)$, depending on an externally tunable parameter $g$, and initially prepared in a thermal state at temperature $T=1/\beta$.
Thus, the system's initial state is given by $\rho_0^\text{th}=e^{-\beta H_0}/Z_0$, where $H_0=H(g_0)$ is the initial Hamiltonian and $Z_0=\tr{e^{-\beta H_0}}$ is the partition function.
The system then undergoes a sudden quench work protocol, where $g$ is changed to a final value $g_\tau$.
Since the quench is assumed to be instantaneous, the state of the system remains the same $\rho_0^\text{th}$, but its Hamiltonian changes to $H_\tau=H(g_\tau)$.
Therefore, the system is driven away from equilibrium.
The entropy production (or nonequilibrium lag) due to this process is given by~\cite{Kawai2007,Landi2020a} 
\begin{equation}\label{entSigma}
    \Sigma = S(\rho_0^\text{th}||\rho_\tau^\text{th}),
\end{equation}
where $S(\rho||\sigma)=\tr{\rho(\ln \rho - \ln \sigma)}\geqslant0$ is the quantum relative entropy, and $\rho_\tau^\text{th}=e^{-\beta H_\tau}/Z_\tau$ is the equilibrium reference state associated with the final Hamiltonian $H_\tau$.
The entropy production $\Sigma$ may also be written in the enlightening form
$\Sigma = \beta (\langle W\rangle - \Delta F)$, where $\langle W\rangle = \tr{(H_\tau - H_0)\rho_0^\text{th}} = \tr{\Delta H \rho_0^\text{th}}$ is the average work performed in the protocol and $\Delta F = F(g_\tau) - F(g_0) = -T \ln Z_\tau/Z_0$ is the difference in equilibrium free energy.
This clear thermodynamic interpretation makes~\eqref{entSigma} an extensively used quantifier of irreversibility in both theory~\cite{Jarzynski1997,Derrida1998,Crooks1998,Kurchan1998,Lebowitz1999,Mukamel2003,Talkner2007,Deffner2010,Guarnieri2018} and experiment~\cite{Liphardt2002,Douarche2005,Collin2005,Speck2007,Saira2012,Koski2013,Batalhao2014,An2014,Batalhao2015,Talarico2016,Zhang2018a,Smith2017}.

Let us consider the quench $\delta g = g_\tau - g_0$ to be small. 
Then we can make a Taylor expansion, and the entropy production can be simplified to~\cite{Varizi2021,Scandi2019}
\begin{IEEEeqnarray}{rCl}
\label{Sigma_Expand}
    \Sigma &=& 
    \sigB + \sigA,
    \\[0.2cm]
\label{B_Expand}
    \sigB &=& \frac{\beta^2}{2}  \, \text{Var}_0 [ \Delta H^\text{d} ],
    \\[0.2cm]
\label{A_Expand}
    \sigA &=& \frac{\beta^2 }{2} \, \text{Var}_0 [ \Delta H^\text{c} ] - \frac{\beta^2}{2} \int_0^1\mathrm{d}y \, I^y(\rho^\text{th}_0,\Delta H^\text{c} ) ,
\end{IEEEeqnarray}
where $\Delta H^\text{d}$ is the diagonal part of the perturbation in the basis of $H_0$ (and $\rho_0^\text{th}$), $\Delta H^\text{c} = \Delta H - \Delta H^\text{d}$ is the coherent part, $\text{Var}_0[\bullet] = \tr{(\bullet)^2\rho_0^\text{th}} - \tr{(\bullet) \rho_0^\text{th}}^2$ is the variance of $(\bullet)$ in the initial thermal state, and
\begin{equation}\label{WYD-info}
    I^y(\varrho,X) = -\frac{1}{2} \tr{ [\varrho^y,X] [\varrho^{1-y}, X] }
\end{equation}
is the Wigner-Yanase-Dyson skew information~\cite{Wigner1963,Wehrl1978}.
The term $\sigA$ quantifies the entropy production associated with the coherences generated by a noncommuting drive.
This expansion works in the range of parameters such that $\beta|\delta g| \lesssim 1$~\cite{Varizi2021}.
And both contributions in Eq.~\eqref{Sigma_Expand} are individually non-negative~\cite{Varizi2021}.

Moreover, for sufficiently high temperatures, the splitting~\eqref{Sigma_Expand} coincides with an alternative expansion used in~\cite{Santos2019,Francica2019}, which reads 
\begin{equation}\label{altsplitting}
    \Sigma =  S(\mathbb{D}_{H_\tau}(\rho_0^\text{th})||\rho_\tau^\text{th}) + S(\rho_0^\text{th}||\mathbb{D}_{H_\tau}(\rho_0^\text{th})),
\end{equation}
where $\mathbb{D}_{H_\tau}(\rho_0^\text{th})$ is the initial state dephased in the final energy basis.
Therefore, the first term gives a contribution due to the mismatch between the populations of the initial and final equilibrium states, while the second gives the contribution to the entropy production stemming from the coherences in $\rho_0^\text{th}$ in the final energy basis as measured by the relative entropy of coherence~\cite{Baumgratz2014}.
For $T$ high  enough, they coincide with $\sigB$ and $\sigA$~\cite{Varizi2021}, respectively.
This means that, in this particular regime of infinitesimal quenches and high temperatures, the splitting~\eqref{Sigma_Expand} and~\eqref{altsplitting} are equivalent.

In what follows, in Sec.~\ref{sec:QCSHT} we will show why and how $\sigB$ and $\sigA$ can be used to investigate a quantum critical point.
Notably, we will see that this approach is useful at any temperature.
In Sec.~\ref{sec:LZ} and~\ref{sec:XYmodel} we consider the Landau-Zener and XY models as examples of quantum critical systems to illustrate our results.
In Sec.~\ref{sec:experiment} we discuss how, in principle, these results could be verified experimentally.
We conclude in Sec.~\ref{sec:conclusion}.

\section{\label{sec:QCSHT}Quantum Critical Signatures at High Temperatures}

We consider a system described by the Hamiltonian
\begin{equation}
    H(g) = H_0 + g H_1,
\end{equation}
where $g$ is an externally adjustable parameter.
If the two parts of this Hamiltonian do not commute, $[H_0,H_1]\neq0$, in the thermodynamic limit the quantum fluctuations induced by $H_1$ as $|g|$ is raised above zero may cause a continuous (second order) quantum phase transition in the system at some critical value $g_c$.
As noted earlier, the existence of this critical point may imprint a signature in physically observable quantities, even if the system is finite in size and/or is in a finite temperature.

Let us assume that $H(g)$ has the following eigendecomposition
\begin{equation}
    H(g) = \sum_i \epsilon_i(g)\, \Pi_i(g),
\end{equation}
where $\epsilon_i$ and $\Pi_i$ are the energy eigenvalues and eigenprojectors, respectively.
These are generally functions of $g$, and differentiating with respect to it, we get
\begin{equation}
    (\partial_g H)\delta g = \delta g \sum_i (\partial_g \epsilon_i)\Pi_i + \delta g\sum_i \epsilon_i(\partial_g \Pi_i).
\end{equation}

On the other hand, suppose we apply an instantaneous perturbation $\delta g$ on the system. 
We have $H(g+\delta g) = H(g) + \Delta H$, where $\Delta H$ can be split as
\begin{equation}
    \Delta H = \Delta H^\text{d} + \Delta H^\text{c},
\end{equation}
with $\Delta H^\text{d} = \sum_i \Pi_i\Delta H\Pi_i$ being the diagonal part of the perturbation in the eigenbasis of the original Hamiltonian $H(g)$, and $\Delta H^\text{c} = \Delta H - \Delta H^\text{d}$ the coherent part.
If we now Taylor expand $H(g+\delta g)$, because $H$ is linear on $g$, we get
\begin{equation}
    \Delta H = H(g + \delta g) - H(g) = (\partial_g H)\delta g,
\end{equation}
from which we readily find that
\begin{IEEEeqnarray}{rCl}
    \Delta H^\text{d} &=& 
    \delta g \sum_i (\partial_g \epsilon_i)\Pi_i,
    \\[0.2cm]
    \Delta H^\text{c} &=& \delta g \sum_i \epsilon_i (\partial_g \Pi_i). 
\end{IEEEeqnarray}

Now consider we prepare the system in thermal equilibrium, as discussed in Sec.~\ref{sec:int}.
For sufficiently high temperatures (small $\beta$), the thermal state $\rho_0^\text{th}=e^{-\beta H(g_0)}/Z(g_0)$ is close to the maximally mixed state, $\rho_0^\text{th} = \maxmix + \mathcal{O}(\beta)$, where $d$ is the dimension of the system.
Further assuming $\delta g$ is small and $\tr{\Delta H^\text{d}}=0$, which can always be done, it is easy to show that to leading order in $\beta$,  Eqs.~\eqref{B_Expand} and~\eqref{A_Expand} reduce to,
\begin{IEEEeqnarray}{rCl}
\label{BInfTemp}
    \sigB &=& 
    \frac{\beta^2\delta g^2}{2}  \sum_i \frac{d_i}{d} (\partial_g \epsilon_i)^2,
    \\[0.2cm]
\label{AInfTemp}
    \sigA &=& 
    \frac{\beta^2\delta g^2}{2}  \sum_i \frac{||\epsilon_i (\partial_g \Pi_i)||^2}{d},
\end{IEEEeqnarray}
where $d_i=\tr{\Pi_i}$ is the dimension of projector $\Pi_i$, $||X|| = \sqrt{\tr{X^\dagger X}}$ is the Hilbert-Schmidt norm of $X$, and we also used that $\Delta H^\text{c}$ is traceless, by construction.
Corrections to these formulas will be of order $\beta^3$ at least, and can be safely ignored for small $\beta$.

From these results, we therefore see that $\sigB$ and $\sigA$ are directly connected to the derivatives of the energy eigenvalues and eigenbasis with respect to the critical parameter of the Hamiltonian.
But by definition~\cite{Sachdev2011}, in the thermodynamic limit, the ground state and ground state energy of a quantum critical system are nonanalytic in the vicinity of a critical point.
Because of the form of Eqs.~\eqref{BInfTemp} and~\eqref{AInfTemp}, these singularities get imprinted on $\sigB$ and $\sigA$ no matter how high the temperature is.
Furthermore, since the thermal state $\rho_0^\text{th}$ is an analytic function of $\beta$, the distinct behaviors of $\sigB$ and $\sigA$ in the vicinity of the critical point $g_c$ persists at every temperature.

Contrastingly, in the limit $T\to\infty$ ($\beta\to0$), using also that $\tr{\Delta H}=\tr{\Delta H^\text{d}}=0$, the total entropy production is simply given by
\begin{equation}
    \Sigma = \frac{\beta^2}{2}\tr{\frac{\mathds{1}}{d}\Delta H^2}.
\end{equation}
But $\Delta H = \delta g(\partial_g H) = \delta g H_1$, which makes $\Sigma$
independent of the critical parameter $g$. This is why in the high-temperature limit, the full entropy production can show no singularity at $g_c$.

Hence, the classical and quantum contributions to the entropy production Eqs.~\eqref{B_Expand} and~\eqref{A_Expand}
can be used to investigate the existence of a second-order quantum critical point
at any temperature.
We now illustrate this via two analytically soluble examples.

\section{\label{sec:LZ}Landau-Zener model}

\begin{figure}
    \includegraphics[width=\linewidth]{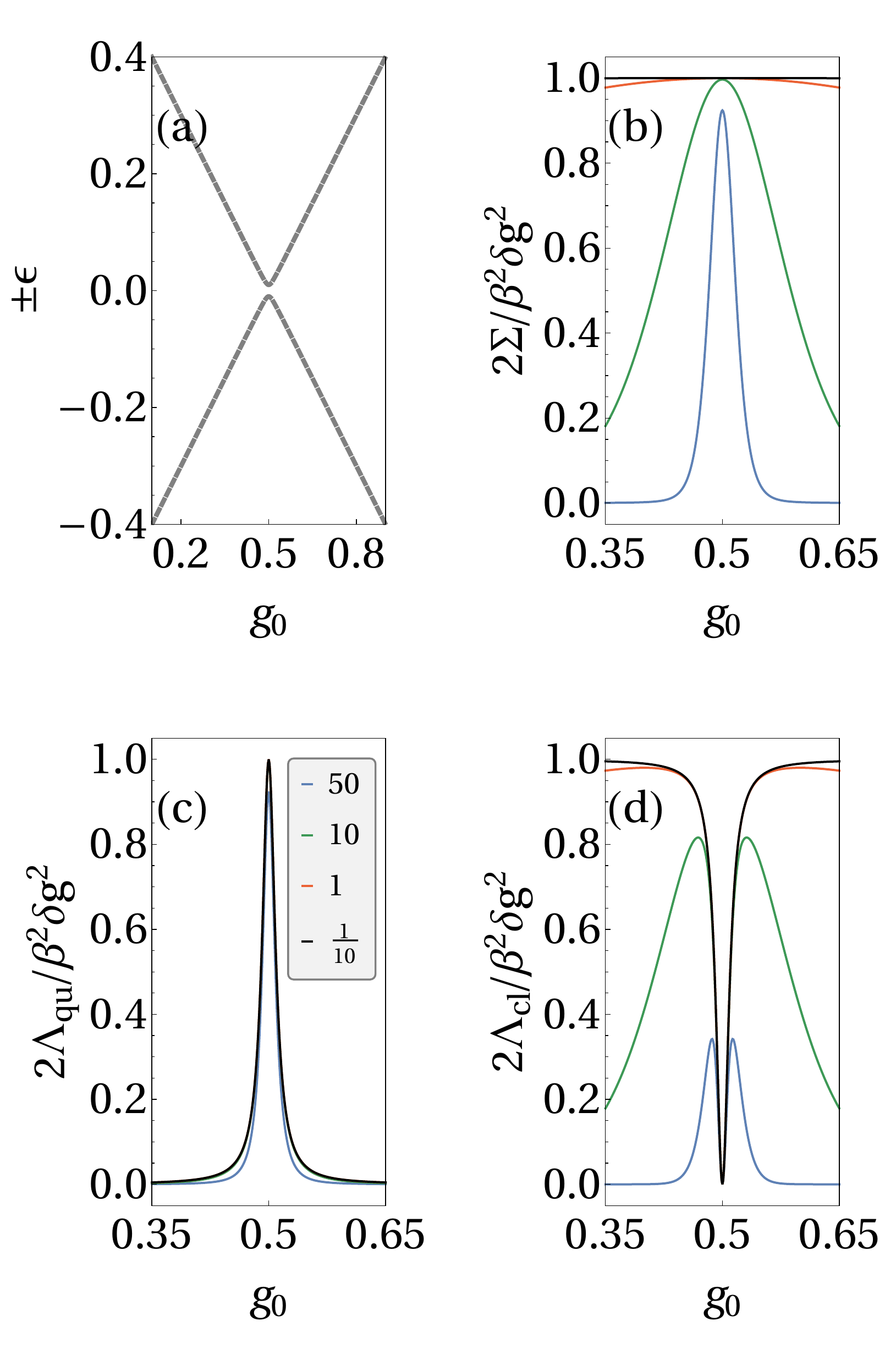}
    \caption{Landau-Zener model. (a) $\pm \epsilon$~vs.~$g_0$,  showing the avoided crossing  at the critical point $g_0 = 1/2$. 
    (b)-(d)  $\Sigma$,  $\sigA$ and $\sigB$ (scaled by $\nicefrac{1}{2}\beta^2 \delta g^2$), for several values of inverse temperatures $\beta$, as denoted in image (c). 
    In $\Sigma$, signatures of the transition are only present at low temperatures (high $\beta$), and are completely washed away for small $\beta$. Conversely, in both $\sigB$ and $\sigA$, clear signatures remain visible over the entire temperature ranges.
    Other parameters: $\Delta=1$ and $b=0.01$.
    Note that since the curves are divided by $\delta g^2$, we do not need to specify a value for it, just assume it is small enough for Eqs.~\eqref{B_Expand} and~\eqref{A_Expand} to be valid.}
    \label{fig:ABLZGrid}
\end{figure}

The Landau-Zener model is a single qubit Hamiltonian which can be regarded as a prototype of a quantum critical system~\cite{Mascarenhas2014}.
It reads
\begin{equation}
    H_\text{LZ} (g) = \Big ( -\frac{\Delta}{2} + g \Big)\sigma^z + b\sigma^x,
\end{equation}
where $\sigma^{x,z}$ are Pauli spin-$\frac{1}{2}$ operators and $g$ represents an externally controlled magnetic field. This system has an avoided crossing at $g_\text{c} = \Delta/2$ for $b\to0$, similar to what happens in a system presenting a second-order quantum phase transition - see Fig.~\ref{fig:ABLZGrid}(a).

The Landau-Zener Hamiltonian assumes the following diagonal form
\begin{equation}
    H_\text{LZ}(g) = \epsilon(g) \tilde{\sigma}^z,
\end{equation}
where $\epsilon(g) = \sqrt{b^2 + (g - \Delta/2)^2}$ gives the system eigenergies, and
\begin{IEEEeqnarray}{rCl}
    \tilde{\sigma}^z &=& |\psi_+\rangle \langle \psi_+| - |\psi_-\rangle \langle \psi_-|,
    \\[0.2cm]
    |\psi_-(g)\rangle &=& \cos(\theta/2)|0\rangle - \sin(\theta/2)|1\rangle,
    \\[0.2cm]
    |\psi_+(g)\rangle &=& \sin(\theta/2)|0\rangle + \cos(\theta/2)|1\rangle,
    \\[0.2cm]
    \big(\cos\theta,\, \sin\theta \big) &=& \bigg(\frac{g - \Delta/2}{\epsilon
    },\, \frac{b}{\epsilon} \bigg).
\end{IEEEeqnarray}
Here $\sigma^z|i\rangle=(-1)^{i+1}|i\rangle$, $i=0,1$, is the usual computational basis.

For a perturbation $\Delta H = \delta g\sigma^z$, the dephased and coherent parts in the energy eigenbasis read
\begin{alignat}{2}
    \Delta H^\text{d} &= \delta g (\partial_g \epsilon) \tilde{\sigma}^z &&= \delta g \cos \theta\, \tilde{\sigma}^z,
    \\[0.2cm]
    \Delta H^\text{c} &= \delta g \epsilon (\partial_g \tilde{\sigma}^z) &&= -\delta g \sin \theta\, \tilde{\sigma}^x,
\end{alignat}
where $\tilde{\sigma}^x = |\psi_-\rangle\langle\psi_+|+|\psi_+\rangle\langle\psi_-|$.
The derivatives $\partial_g\epsilon$ and $\partial_g\tilde{\sigma}^z$ are discontinuous at $g_c$ when $b\to0$, and this is reflected in the behaviors of $\sigB$ and $\sigA$ around this point.

Consider a system initially in the thermal state
\begin{equation}
    \rho_0^\text{th} = \frac{e^{-\beta H_\text{LZ}(g_0)}}{Z_0} = \frac{e^{ -\beta \epsilon^0} |\psi_+^0\rangle \langle\psi_+^0| + e^{\beta \epsilon^0} |\psi_-^0\rangle \langle\psi_-^0| }{2\cosh(\beta\epsilon^0)},
\end{equation}
where $\epsilon^0=\epsilon(g_0)$ and $|\psi_{\pm}^0\rangle=|\psi_{\pm}(g_0)\rangle$.
Hence, for small $\delta g$, using Eqs.~\eqref{B_Expand} and~\eqref{A_Expand}, we have
\begin{IEEEeqnarray}{rCl}
    \sigB &=& \frac{1}{2} \beta^2 \delta g^2 \sech^2(\beta \epsilon^0) \cos^2\theta,
    \\[0.2cm]
    \sigA &=& \frac{1}{2} \beta^2 \delta g^2 \frac{\tanh(\beta \epsilon^0)}{\beta \epsilon^0} \sin^2\theta,
\end{IEEEeqnarray}
which reduce to $\sigB = (1/2)\beta^2\delta g^2 \cos^2\theta$ and $\sigA = (1/2)\beta^2\delta g^2\sin^2\theta$ to leading order on $\beta$, consistent with Eqs.~\eqref{BInfTemp} and~\eqref{AInfTemp}.
In Fig.~\ref{fig:ABLZGrid}(b-d) we plot $\Sigma$, $\sigA$ and $\sigB$ as a function of the initial field for several inverse temperatures $\beta$.
The net entropy production $\Sigma$ shows signatures of the transition at low temperatures. But these are quickly washed away and for high $T$ (small $\beta$), $\Sigma$ becomes essentially flat. 
Conversely, the curves for $\sigA$ and $\sigB$ preserve the signatures of the transition for all values of $\beta$, indicated by a sharp peak (dip) in the plots of $\sigA$ ($\sigB$) at $g_c$.

\section{\label{sec:XYmodel}Transverse field XY model}

\begin{figure*}
    \centering
    \includegraphics[width=\textwidth]{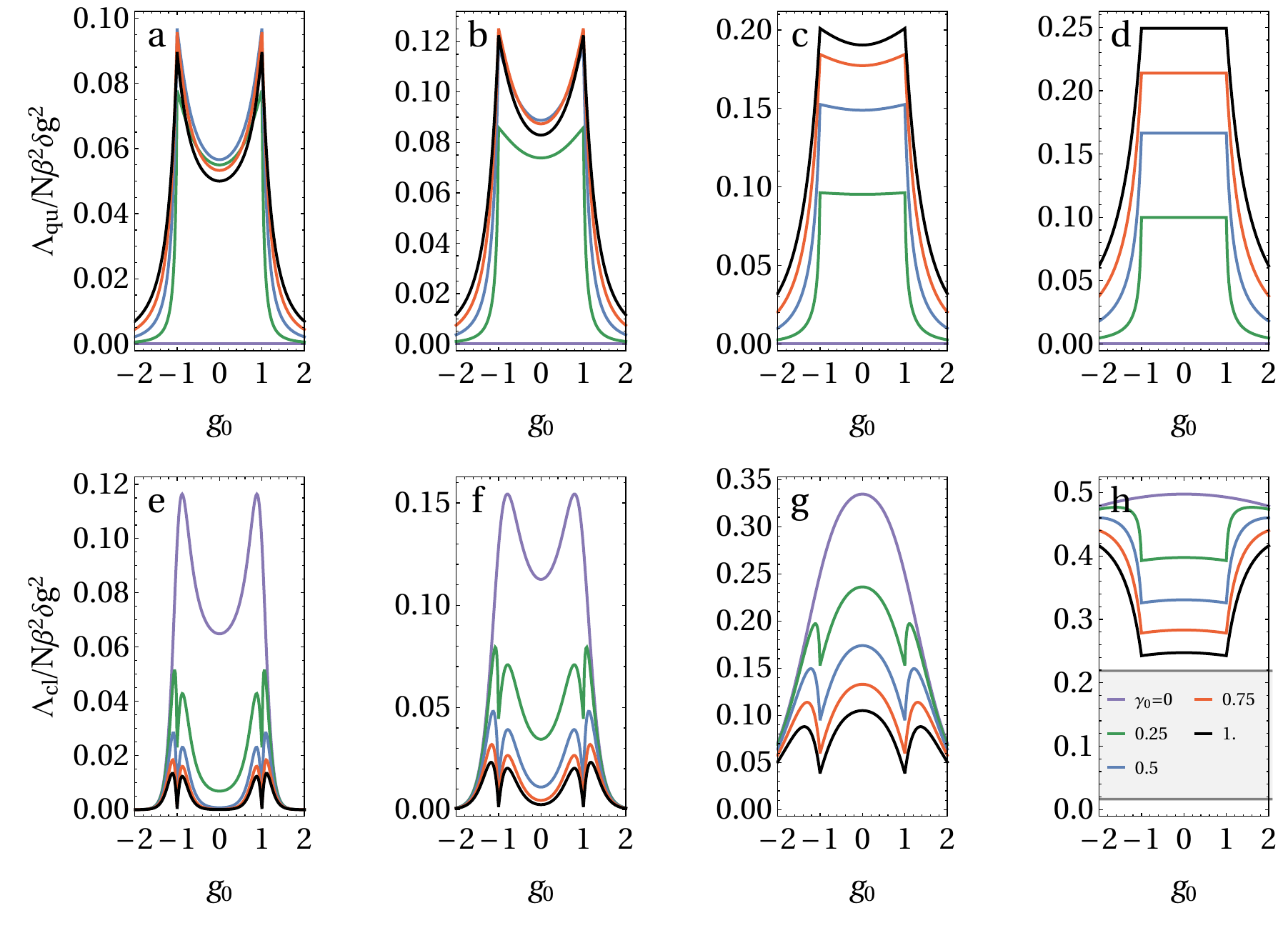}
    \caption{\label{fig:IsingTransitionGrid} Plots of $\sigA$ (above) and $\sigB$ (below) as a function of the initial field $g_0$, scaled by $N\beta^2\delta g^2$, for several values of anisotropy and inverse temperatures (a)/(e) $\beta = 5$, (b)/(f) $\beta = 3$, (c)/(g) $\beta=1$ and (d)/(h) $\beta=0.1$.
    We consider the anisotropy to be fixed, $\delta\gamma =0$, and make small quenches $\delta g$ in the field.
    Note that divided by $\delta g^2$, the curves are independent of its value.
    For $\gamma=0$, the system's eigenbasis is constant and, therefore, $\sigA=0$.
    For other values of $\gamma$, a kink in both quantities clearly indicates the critical points $g_c = \pm 1$, even at high temperatures.
    Particularly, in the limit $T\to\infty$ ($\beta\to0$) we obtain the plateaus/depressions in panel (d)/(h).}
\end{figure*}
\begin{figure*}
    \centering
    \includegraphics[width=\textwidth]{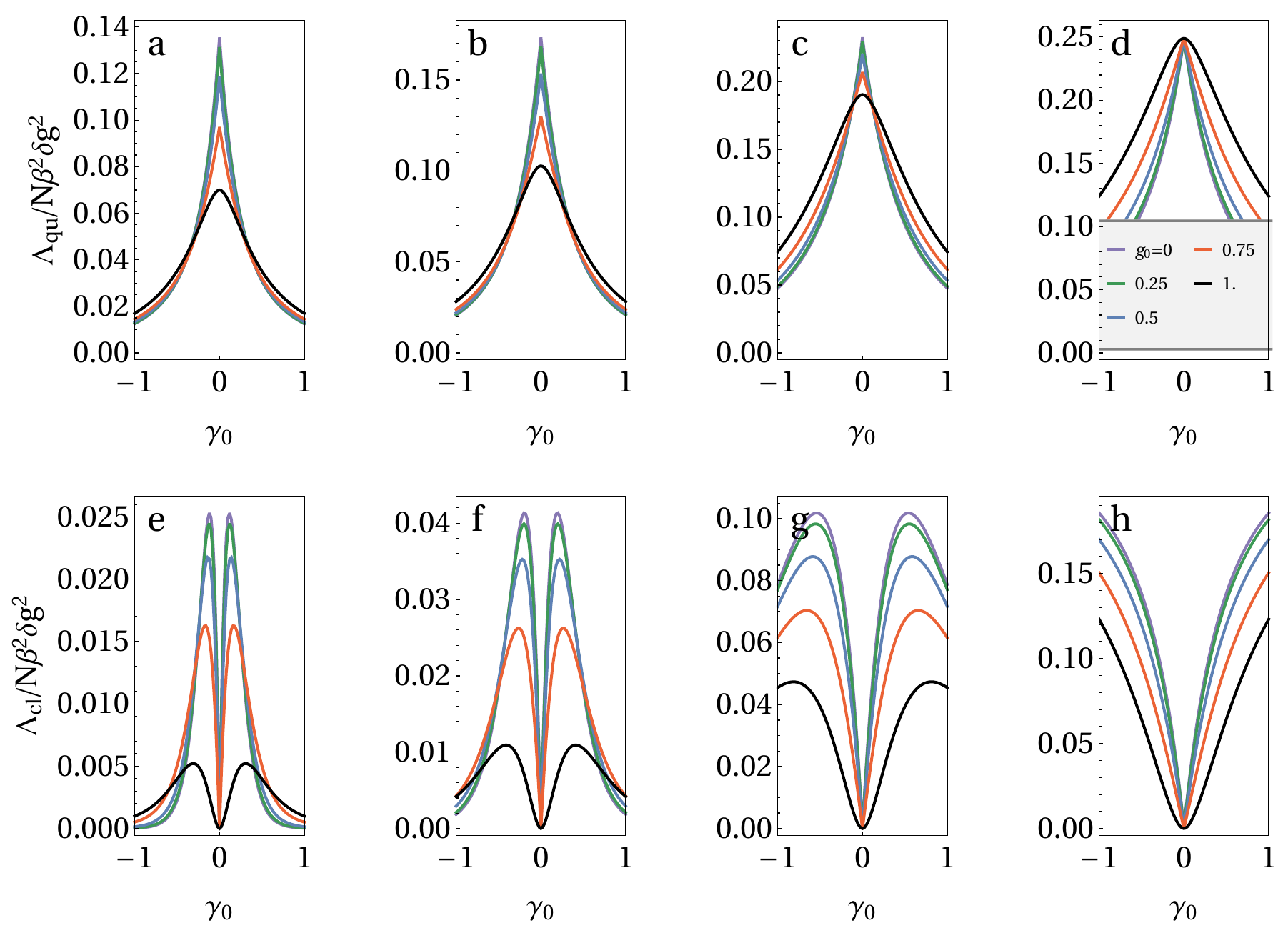}
    \caption{\label{fig:AnisotropicTransitionGrid} Analagous to Fig.~\ref{fig:IsingTransitionGrid} but analysing the anisotropic transition.
    Hence the quenches are made on the anisotropic parameter, with $\delta g=0$.
    Divided by $\delta\gamma^2$, the curves are independent of its value.
    The peaks in $\sigA$ above (dips in $\sigB$ below) evidence the quantum critical line at $\gamma=0$.
    Again, each column has an inverse temperature which is equal to the corresponding column in Fig.~\ref{fig:IsingTransitionGrid}.}
\end{figure*}

Next, we turn to the transverse field XY model, described by a linear chain with $N$ spins, which has the Hamiltonian 
\begin{equation}\label{XYHamiltonian}
    H(g,\gamma) = -J\sum_{j=1}^N \bigg( \frac{1+\gamma}{2}\sigma_j^x\sigma_{j+1}^x + \frac{1-\gamma}{2}\sigma_j^y\sigma_{j+1}^y + g\sigma_j^z \bigg),
\end{equation}
where $J>0$ is the ferromagnetic exchange interaction between spins, $g$ is an applied magnetic field and $\gamma$ is the anisotropy parameter.
We also consider periodic boundary conditions, $\vec{\sigma}_{N+1}=\vec{\sigma}_1$.
For $\gamma=1$,~\eqref{XYHamiltonian} reduces to the transverse field Ising model, and $\gamma=0$ gives the $XX$ model.
In the latter, the system eigenbasis is constant (does not depend on $g$).
Without loss of generality we can make $J=1$.

Perturbations on the system may be introduced by varying $g$ and/or $\gamma$.
In the thermodynamic limit, this model presents critical lines at $g=\pm1$, where the system changes from a ferromagnetic phase, $|g|<1$, to a paramagnetic phase for $|g|>1$.
There is also an anisotropic transition line at $(\gamma=0,|g|<1)$ where the ferromagnetic ordering changes from the $y$-direction, for $\gamma<0$, to the $x$-direction, for $\gamma>0$.

After a Jordan-Wigner transformation that maps the system onto spinless fermions, and a Fourier and Bogoliubov transformations, the Hamiltonian~\eqref{XYHamiltonian} can be written in diagonal form as (we ignore parity issues for simplicity and assume $N$ is even)~\cite{Sachdev2011,Damski2014}
\begin{equation}
    H(g,\gamma) = \sum_k \epsilon_k\, (2\eta_k^\dagger\eta_k -1),
\end{equation}
where $k=\pm(2n+1)\pi/N$, $n=0,1,...,N/2-1$ are the system quasi-momenta; $\epsilon_k(g,\gamma) = \sqrt{(g - \cos k)^2 + \gamma^2\sin^2 k}$ are the single particle eingenergies; and
\begin{IEEEeqnarray}{rCl}
    \label{BogTrans}
    \eta_k &=& \cos(\theta_k/2) c_k + \sin(\theta_k/2) c_{-k}^{\dagger},
    \\[0.2cm]
    \big(\cos\theta_k, \sin\theta_k\big) &=& \Big(\frac{g - \cos k}{\epsilon_k(g,\gamma)}, \frac{\gamma\sin k}{\epsilon_k(g,\gamma)}\Big),
\end{IEEEeqnarray}
where $\{\eta_k\}$ and $\{c_k\}$ are fermionic operators, satisfying the usual anticommutation relations~\cite{Sachdev2011,Varizi2020}.
Note that the set  $\{c_k\}$ is independent of $g$ and $\gamma$.

For a perturbation $\Delta H = - \delta \gamma/2 \sum_{j=1}^N(\sigma_j^x\sigma_{j+1}^x -\sigma_j^y\sigma_{j+1}^y) - \delta g \sum_{j=1}^N \sigma_j^{z}$,  its dephased and coherent parts were shown in Ref.~\cite{Varizi2021} to be given by
\begin{IEEEeqnarray}{rCl}
    \Delta H^\text{d} &=& \sum_{k}  (\delta \gamma \sin k \sin \theta_k + \delta g\cos\theta_k) (2\eta_k^{\dagger}\eta_k - 1),
    \IEEEeqnarraynumspace\\[0.2cm]
    \Delta H^\text{c} &=& \sum_k (\delta g\sin\theta_k - \delta \gamma \sin k \cos \theta_k)  (\eta_{-k}^{\dagger}\eta_k^{\dagger} - \eta_{-k}\eta_k).\IEEEeqnarraynumspace
\end{IEEEeqnarray}

Considering the initial thermal state $\rho_0^\text{th}=e^{-\beta H_0}/Z_0$, with $H_0=H(g_0,\gamma_0)$, performing small sudden quenches $\delta g$ and $\delta\gamma$, we obtain
\begin{IEEEeqnarray}{rCl}
\label{XY_final_sigB}
    \frac{\sigB}{N\beta^2} &=& \int_0^\pi \frac{\mathrm{d}k}{2\pi} \sech^2(\beta \epsilon_k^0)(\delta \gamma \sin k \sin \theta_k + \delta g\cos\theta_k)^2,\IEEEeqnarraynumspace
    \\[0.2cm]
\label{XY_final_sigA}    
    \frac{\sigA}{N\beta^2} &=& \int_0^\pi \frac{\mathrm{d}k}{2\pi}\frac{\tanh(\beta \epsilon_k^0)}{\beta\epsilon_k^0} (\delta g\sin\theta_k - \delta \gamma \sin k \cos \theta_k)^2,\IEEEeqnarraynumspace
\end{IEEEeqnarray}
where we took the thermodynamic limit $N\to\infty$ to convert the sums over $k$ into integrals, and $\epsilon_k^0=\epsilon_k(g_0,\gamma_0)$ are the eigenenergies associated with the initial values of anisotropy and field.

Eqs.~\eqref{XY_final_sigB} and~\eqref{XY_final_sigA} are general expressions for the quench response of $\sigB$ and $\sigA$, valid for any value of $\beta$. 
In Fig.~\ref{fig:IsingTransitionGrid} we plot them assuming small quenches $\delta g$ in the field, with $\delta \gamma=0$. Each curve correspond to a value of the anisotropy $\gamma$, and each column to a value of inverse temperature: (a)/(e) $\beta=5$, (b)/(f) $\beta=3$, (c)/(g) $\beta=1$ and (d)/(h) $\beta=0.1$.
For $\gamma=0$, the system's constant eigenbasis makes $\sigA=0$.
In all other curves, there are kinks at $g=\pm1$, indicating the critical points at high temperatures.
In particular, for sufficiently small $\beta$ we have
\begin{IEEEeqnarray}{rCl}
\label{B-smallbeta}
    \frac{\sigB}{N\beta^2\delta g^2} &=& \int_0^{\pi}\frac{\mathrm{d}k}{2\pi}\cos^2\theta_k,
    \\[0.2cm]
\label{A-smallbeta}
    \frac{\sigA}{N\beta^2\delta g^2} &=& \int_0^{\pi}\frac{\mathrm{d}k}{2\pi}\sin^2\theta_k,
\end{IEEEeqnarray}
which gives a total entropy production $\Sigma/N\beta^2\delta g^2=1/2$, containing no distinct feature whatsoever.
Nonetheless, for $\gamma=1$, the integral in~\eqref{A-smallbeta} evaluates to~\cite{Varizi2020}
\begin{equation}\label{A-smallbeta-Ising}
     \frac{\sigA}{N\beta^2\delta g^2} = \begin{cases}
     \frac{1}{4},\quad\, |g_0|\leqslant1,\\
     \frac{1}{4|g_0|},\, |g_0|>1,
     \end{cases}
\end{equation}
corresponding to the striking plateaus in the region of parameters associated with the quantum ferromagnetic phase observed in Fig.~\eqref{fig:IsingTransitionGrid}(d)/(h).
Note that these curves are for $\beta=0.1$, corresponding to a large temperature.
In Appendix~\ref{appsec:finiteN} we also analyse the effects of a finite number of spins $N$, and show that although the functions become analytic, the thermodynamic limit behavior is quickly approached with increasing $N$.
This mean our analysis could be tested in an Ising model with a relatively small number of spins.

In Fig.~\ref{fig:AnisotropicTransitionGrid} we  analyse the anisotropic transition by taking $\delta g=0$, and a small $\delta \gamma$.
In this case, each curve corresponds to a value of transverse field $g_0$, and, again, each column to an inverse temperature, with the same values as those in Fig.~\ref{fig:IsingTransitionGrid}.
A cusp in $\sigB$ and $\sigA$ at $\gamma_0=0$ signal the existence of the quantum critical line associated with the change in the ferromagnetic ordering, even at high temperatures.


\section{\label{sec:experiment} Experimental Assessment of $\sigB$ and $\sigA$}

In this section we propose a way in which the classical and quantum contributions to the entropy production, $\sigB$ and $\sigA$, could, in principle, be evaluated experimentally.
Our idea is based on their stochastic formulation using the standard two-point measurement scheme~\cite{Talkner2007} and the fact that they obey fluctuation theorems~\cite{Varizi2021}.

The system is initially prepared in the thermal state $\rho_0^\text{th}=e^{-\beta H_0}/Z_0=(e^{-\beta \epsilon_i^0}/Z_0) |i_0\rangle \langle i_0|$ at inverse temperature $\beta$, associated with the Hamiltonian $H_0 = \sum_i \epsilon_i^0|i_0\rangle\langle i_0|$, which we assume nondegenerate, for simplicity.
Hence, if we perform an energy measurement in this state, we obtain the energy $\epsilon_i^0$ with probability $p_i^0 = e^{-\beta \epsilon_i^0}/Z_0$, while the associated measurement backaction updates the state of the system to $|i_0\rangle$.
We then perform the quench, and the Hamiltonian changes to the final value $H_\tau = \sum_j \epsilon_j^\tau |j_\tau\rangle \langle j_\tau|$. 
Since we are considering an instantaneous quench, and given that after the first energy measurement, the system is in the state $|i_0\rangle$, a second energy measurement will return the value $\epsilon_j^\tau$ with probability $p_{i\,|j}=|\langle j_\tau|i_0\rangle|^2$.

Therefore, the path probability corresponding to the stochastic trajectory $|i_0\rangle \to |j_\tau\rangle$ is given by the product of the probabilities for the system to be initially found in $|i_0\rangle$ and to transition to the final state $|j_\tau\rangle$, after the quench;
it reads $P_F[i,j]=p_i^0 p_{i\,|j}$.
The associated stochastic entropy production is given by~\cite{Francica2019}
\begin{equation}
    \sigma[i,j]= \beta (\epsilon_j^\tau - \epsilon_i^0) - \beta \Delta F_{\tau, \, 0},
\end{equation}
where $w[i,j]=\epsilon_j^\tau - \epsilon_i^0$ is the stochastic work done on the system, and $\Delta F_{\tau,0}=-T\ln Z_\tau/Z_0$ is the change in equilibrium free energy.

It can be readily checked that $\langle \sigma[i,j] \rangle = \sum_{i,j}\sigma[i,j]P_F[i,j]=\Sigma$ gives the expected average~\eqref{entSigma}.
Moreover, the quantity $\sigma$ satisfies an integral fluctuation theorem~\cite{Esposito2009,Campisi2011}, $\langle e^{-\sigma}\rangle = 1$, from which follows the Jarzynski relation $\langle e^{-\beta w[i,j]}\rangle = e^{-\beta \Delta F_{\tau,0}}$~\cite{Jarzynski1997}.

Thus, we see that if in an experiment, one can determine the work $w[i,j]$ and the path probability $P_F[i,j]$, $\Delta F_{\tau,0}$ is obtained from the Jarzynski relation and so, finally, the entropy production $\Sigma$.
There are several proposals on how this can be done, as well as successful experimental implementations~\cite{Dorner2013,Mazzola2013a,Roncaglia2014a,Batalhao2014,An2014,Cerisola2017}.
In~\cite{Herrera2021} the authors present a general method for obtaining the transition probabilities $p_{i\,|j}$ in a many-body system and use it to verify the Jarzynski relation for two qubits.

Having said that, next we note that $\sigB$ and $\sigA$ have similar stochastic versions~\cite{Varizi2021},
\begin{IEEEeqnarray}{rCl}
        \label{stoch-b}
    \sigBij[i,j] &=& \beta (\tilde{\epsilon}_i^\tau - \epsilon_i^0) - \beta \Delta \tilde{F}_{\tau, \, 0},
    \\[0.2cm]
    \sigAij[i,j] &=& \beta (\epsilon_j^\tau - \tilde{\epsilon}_i^\tau) - \beta \Delta \tilde{F}_{\tau,\,\tau},
\end{IEEEeqnarray}
where $\tilde{\epsilon}_i^\tau = \epsilon_i^0 + \Delta H_{ii} = \epsilon_i^0 + \delta g (\partial_g\epsilon_i^0)$ are the eigenenergies associated with the Hamiltonian $H_0 + \Delta H^\text{d}$, and $\Delta \tilde{F}_{\tau,0} = -T\ln \tilde{Z}_\tau/Z_0$, with $\tilde{Z}_\tau = \tr{e^{-\beta(H_0 + \Delta H^\text{d})}}$, is the change in the equilibrium free energy due to the incoherent perturbation $\Delta H^\text{d}$.
Additionally, $\Delta \tilde{F}_{\tau,\tau} = -T\ln Z_{\tau}/\tilde{Z}_{\tau}$, gives the difference in free energy associated with the perturbation's coherent part $\Delta H^\text{c}$.

Naturally, we have $\langle \sigBij\rangle = \sum_{i,j}\sigBij[i,j]P_F[i,j] = \sigB$ and $\langle \sigAij \rangle = \sum_{i,j}\sigAij[i,j]P_F[i,j] = \sigA$~\cite{Varizi2021}.
Furthermore, $\sigBij$ satisfies the integral fluctuation theorem  $\langle e^{-\sigBij}\rangle = 1$, and this is equally valid for $\sigAij$ in the infinitesimal and instantaneous quench limit~\cite{Varizi2021}.

Now we observe that, up to second order on the perturbation $\delta g$, the final energy eigenvalues are given by
\begin{equation}\begin{aligned}
\label{EnergFinExpan}
    \epsilon_j^\tau &= \epsilon_j^0 + \delta g (\partial_g\epsilon_j^0) + \frac{1}{2} \delta g^2 (\partial_g^2\epsilon_j^0)
    \\[0.2cm]
    &= \epsilon_j^0 + \Delta H_{jj} + \frac{1}{2}\sum_{\ell\neq j} \frac{|\Delta H_{j\ell}|^2}{\epsilon_j^0 - \epsilon_\ell^0},
\end{aligned}\end{equation}
where $\Delta H_{ij}=\langle i_0|\Delta H|j_0\rangle$.

Interestingly, in this limit of instantaneous and infinitesimal quenches, the first term in Eq.~\eqref{stoch-b} $w^\text{d}[i,i] = \tilde{\epsilon}_i^\tau - \epsilon_i^0 \approx w[i,j]\delta_{ij}$, except in the vicinity of a critical point. 
In the latter case, the third term in the r.h.s. of Eq.~\eqref{EnergFinExpan} can become relevant.
Nonetheless, the smaller $\delta g$ is, the better the approximation and smaller the interval where it breaks down. In the limit $\delta g \to 0$, it should fail only at the critical point.
Hence, we can obtain $w^\text{d}$ from the measured stochastic work $w$ by \emph{post-selecting} the cases where the system changes energy without jumping to a different state.
Using again a Jarzynski relation, $\langle e^{-\beta w^\text{d}}\rangle = e^{-\beta \Delta \tilde{F}_{\tau,\,0}}$, we get $\Delta \tilde{F}_{\tau,\,0}$, and $\sigB = \beta \langle w^\text{d}\rangle - \beta \Delta \tilde{F}_{\tau,\,0}$.
From $\sigB$ and $\Sigma$, we also obtain $\sigA = \Sigma - \sigB$.
This approach therefore enables $\sigA$ and $\sigB$ to be determined from the standard two-measurement protocol, provided the data is post-select, which is highly convenient. 


\section{\label{sec:conclusion}Conclusion}

The entropy production associated with the work done on a closed quantum system can be divided into a classical and a quantum contributions~\cite{Santos2019,Francica2019,Scandi2019,Varizi2021}.
The later originates from energetic coherences that can be generated by the drive.

In this article we showed that for instantaneous and infinitesimal quenches, these contributions, as given in Eqs.~\eqref{B_Expand} and~\eqref{A_Expand}, are respectively and explicitly related to the derivatives of the energy eigenvalues and eigenbasis with respect to the work parameter.
In a system presenting a second-order quantum phase transition, one of these energy eigenvalues and eigenstates becomes nonanalytic at a critical point, and this pathology gets engraved on $\sigB$ and $\sigA$.
In particular, we demonstrated that their singularities remain present even when the system is prepared at arbitrarily high temperatures. 
We believe this makes these quantities particularly useful in spotlighting such quantum critical points.

We illustrated this idea by appliying our general results in two paradigmatic examples, the Landau-Zener and $XY$ models.
The Landau-Zener model is a single qubit Hamiltonian incorporating an avoided energy crossing, analogous to a second-order quantum phase transition.
As a function of the initial work parameter $g_0$, the entropy production in this model has a peak at the critical point $g_c$, at low temperatures.
As $T$ is raised (or $\beta=1/T$ is lowered) the relative height (when the curve is divided by $\beta^2$) of this peak decreases, and the curve becomes flat in the limit $T\to\infty$ ($\beta\to0$).
The maximum of $\sigA$ (minimum of $\sigB$) at $g_c$, however, remain present even in this limit.
Therefore, $\sigB$ and $\sigA$ can be used to spot the critical point even when the system initially has a high temperature.

In the case of the $XY$ model we considered both the ferromagnetic and anisotropic transitions.
In the former we considered quenches in the field with fixed anisotropy and showed that in the thermodynamic limit $\sigB$ and $\sigA$ per particle have kinks at initial fields equal to the critical values $g_0=\pm 1$ in all temperatures.
Notably, in the limit $T\to\infty$, the total entropy production becomes a constant function of $g_0$, while $\sigB$ and $\sigA$ present very distinct behaviors in the regions corresponding to different phases, being constant for $|g_0|<1$ but strictly monotonic when $|g_0|>1$.
We also considered the effects of a finite chain for the Ising model in the appendix.
Although $\sigB$ and $\sigA$ become analytic in this case, we see the thermodynamic limit behavior is quickly approached with increasing $N$.

Similarly, considering quenches in the anisotropy, a kink in the curves of $\sigB$ and $\sigA$ at the critical value $\gamma_0=0$ signal the presence of the anisotropic critical line in high temperatures.

Finally, we also suggested a way in which these results could be obtained from an experiment.
This is based on the stochastic two-point measurement definitions of $\sigB$ and $\sigA$ and the fact that they obey fluctuation theorems.
The actual feasibility of our scheme, however, depends on ones ability to determine the stochastic work performed in the protocol and the associated path probabilities.

\section*{Acknowledgements}

We acknowledge financial support from the Brazilian agencies Conselho Nacional de Desenvolvimento Cient\'ifico e Tecnol\'ogico (CNPq) and Coordena\c c\~ao de Aperfei\c coamento de Pessoal de N\'ivel  Superior - Brasil (CAPES) - Finance Code 001.
GTL acknowledges the financial support of the S\~ao Paulo Funding Agency FAPESP (Grant No.~2019/14072-0.), and the Brazilian funding agency CNPq (Grant No. INCT-IQ 246569/2014-0).

\appendix

\section{\label{appsec:finiteN} Finite-size effects in the Ising model}

In this appendix we consider the effects of a finite number of spins $N$ to the behaviors of $\sigB$ and $\sigA$.

In particular we take $\gamma=1$, which means we work with the transverse field Ising model.
We also assume the temperature is high enough so that we can approximate the initial thermal state by the maximally mixed state, $\rho_0^\text{th} = \mathds{1}/2^N$, where $2^N$ is the dimension of the Hilbert space of a system with $N$ spins.
Then, we have
\begin{IEEEeqnarray}{rCl}
        \label{finiteNsigB}
        \sigB &=& \frac{\beta^2}{2}\tr{\frac{\mathds{1}}{2^N}(\Delta H^\text{d})^2} = \beta^2\delta g^2 \sum_{k>0} \cos^2\theta_k,
        \\[0.2cm]
        \label{finiteNsigA}
        \sigA &=& \frac{\beta^2}{2}\tr{\frac{\mathds{1}}{2^N}(\Delta H^\text{c})^2} = \beta^2\delta g^2 \sum_{k>0} \sin^2\theta_k,
\end{IEEEeqnarray}
where $k=\pm(2n+1)\pi/N$, $n$ ranging from $0$ to $N/2-1$, $\cos\theta_k = (g_0 - \cos k)/\epsilon_k^0$ and $\sin\theta_k = \sin k/\epsilon_k^0$, with $\epsilon_k^0 = \sqrt{(g_0 - \cos k)^2 + \sin^2 k}$ and $g_0$ being the initial field.
We also used that $\cos^2\theta_k$ and $\sin^2\theta_k$ are even functions of $k$.

\begin{figure}[h]
    \centering
    \includegraphics[width=\linewidth]{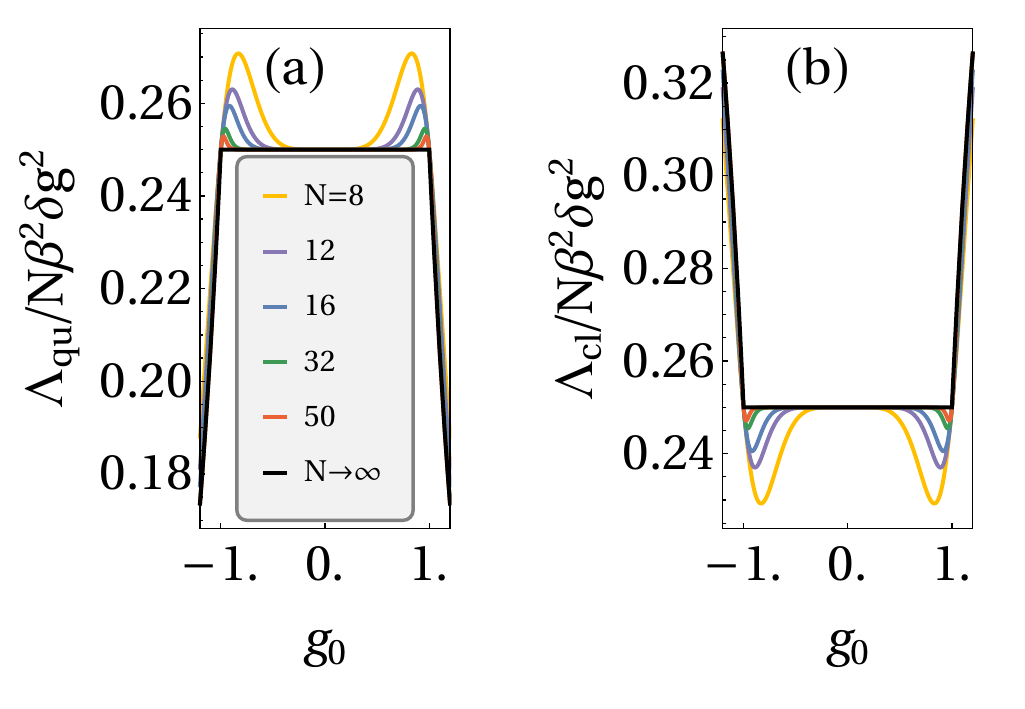}
    \caption{\label{fig:finiteIsing} Plots of (a) $\sigA$ and (b) $\sigB$ scaled by $N\beta^2\delta g^2$ as a function of the initial field $g_0$, for several values of $N$.
    The figure shows the limit $N\to\infty$ is quickly approached with increasing $N$.
    We emphasize that we are considering $\beta$ sufficiently small such that we take only the leading order contributions.}
\end{figure}

In Fig.~\ref{fig:finiteIsing} we plot $\sigB/N$ and $\sigA/N$ as a function of $g_0$ for several $N$.
The plots show that the thermodynamic limit behavior (black curves) is fastly approximated with increasing $N$.

This result can be further elucidated as follows. Eqs.~\eqref{finiteNsigB} and~\eqref{finiteNsigA} can be recast as
\begin{IEEEeqnarray}{rCl}
        \label{finiteNsigB2}
        \frac{\sigB}{N\beta^2\delta g^2} &=& \sum_{k>0} \cos^2\theta_k \frac{\Delta k}{2\pi},
        \\[0.2cm]
        \label{finiteNsigA2}
        \frac{\sigA}{N\beta^2\delta g^2} &=& \sum_{k>0} \sin^2\theta_k \frac{\Delta k}{2\pi},
\end{IEEEeqnarray}
where $\Delta k = 2\pi/N$. 
If we consider a partition
\begin{equation*}
P=\bigg\{\bigg[0,\frac{2\pi}{N} \bigg], \bigg[ \frac{2\pi}{N},\frac{4\pi}{N} \bigg], ..., \bigg[ \pi-\frac{2\pi}{N},\pi \bigg] \bigg\}
\end{equation*}
of the interval $[0,\pi]$, the r.h.s. of~\eqref{finiteNsigB2} and~\eqref{finiteNsigA2} are equivalent to midpoint Riemann sums of the functions $\cos^2\theta_k$ and $\sin^2\theta_k$ over $[0,\pi]$ with partition $P$.
The absolute difference between these sums and the respective thermodynamic limit integrals in~\eqref{B-smallbeta} and~\eqref{A-smallbeta} is bounded by
\begin{equation}
    \Bigg|\int_0^\pi \frac{\mathrm{d}k}{2\pi} f(k) - \sum_{k>0}f(k) \frac{\Delta k}{2\pi} \bigg| \leqslant \frac{M \pi^3}{6 N^2},
\end{equation}
where $f(k)$ is either $\cos^2\theta_k$ or $\sin^2\theta_k$, and $M = \max_{k\in[0,\pi]}|\partial_k^2 f(k)|$ is the maximum absolute value of the second derivative of $f(k)$ in the interval $[0,\pi]$.
Note that $f$ is also a function of the initial field $g_0$, but we omit this for simplicity of notation.
This derivative reads
\begin{equation}\begin{aligned}
    |\partial_k^2 f(k)| = \frac{2}{(\epsilon_k^0)^4} | &\sin^2 \theta_k (g_0^2 - 1)^2 +
    \\[0.2cm]
    &\, \epsilon_k^0 \cos\theta_k(1-g_0\cos k)\cos k|.
\end{aligned}\end{equation}

One can graphically check that the maximum of this function occur at the boundary $k=0\,(\pi)$ for $g_0\geqslant0\,(g_0<0)$.
Specifically, we have
\begin{equation}
    M = \begin{cases}
      \frac{2}{(g_0 + 1)^2},\quad\text{ if } g_0<0,\\
      \frac{2}{(g_0 - 1)^2},\quad\text{ if } g_0\geqslant0.
    \end{cases}
\end{equation}
Hence, away from the critical points, the values of $\sigB/N$ and $\sigA/N$ for a finite chain converge to the thermodynamic limit with a swiftly decreasing error of order $1/N^2$.
The divergence of $M$ at the critical points $g_0=\pm 1$ also clarify the slower convergence in these regions. 


\bibliography{library,bib}
\end{document}